%% file: main.tex
\documentclass[12pt]{article}
\usepackage{styles/mystyle} 
\usepackage{dsfont} 

\begin{document}
\title{Prioritizing Investments in Cybersecurity: Empirical Evidence from an Event Study on the Determinants of Cyberattack Costs}
\date{\today}
\author{Daniel Celeny\thanks{Swiss Finance Institute, EPFL - Cyber-Defence Campus, armasuisse S+T. \href{mailto:daniel.celeny@alumni.epfl.ch}{daniel.celeny@alumni.epfl.ch}},
Loïc Maréchal\thanks{HEC Lausanne, University of Lausanne, \href{mailto:loic.marechal@unil.ch}
{loic.marechal@unil.ch}},
Evgueni Rousselot\thanks{Cyber-Defence Campus, armasuisse S+T, \href{mailto:evgueni.rousselot@epfl.ch}{evgueni.rousselot@epfl.ch}},
Alain Mermoud\thanks{Cyber-Defence Campus, armasuisse S+T, \href{mailto:mermouda@ethz.ch}{mermouda@ethz.ch}}, and
Mathias Humbert\thanks{HEC Lausanne, University of Lausanne, \href{mailto:mathias.humbert@unil.ch}\newline\newline This document results from a research project funded by the Cyber-Defence Campus, armasuisse Science and Technology. Our code will be available at \url{https://github.com/technometrics-lab/17-The-determinants-of-cyberattack-costs-An-event-study} Corresponding author: Daniel Celeny e-mail: \href{daniel.celeny@alumni.epfl.ch}{daniel.celeny@alumni.epfl.ch}, Cyber-Defence Campus, Innovation Park, EPFL, 1015 Lausanne.}}

\renewcommand{\thefootnote}{\fnsymbol{footnote}}
\singlespacing
\maketitle
\vspace{-.2in}


\begin{abstract}


Along with the increasing frequency and severity of cyber incidents, understanding their economic implications is paramount. In this context, listed firms' reactions to cyber incidents are compelling to study since they (i) are a good proxy to estimate the costs borne by other organizations, (ii) have a critical position in the economy, and (iii) have their financial information publicly available.
We extract listed firms' cyber incident dates and characteristics from newswire headlines. We use an event study over 2012--2022, using a three-day window around events and standard benchmarks. We find that the magnitude of abnormal returns around cyber incidents is on par with previous studies using newswire or alternative data to identify cyber incidents. Conversely, as we adjust the standard errors accounting for event-induced variance and residual cross-correlation, we find that the previously claimed significance of abnormal returns vanishes. Given these results, we run a horse race of specifications, in which we test for the marginal effects of type of cyber incidents, target firm sector, periods, and their interactions. Data breaches are the most detrimental incident type with an average loss of -1.3\% or (USD -1.9 billion) over the last decade. The health sector is the most sensitive to cyber incidents, with an average loss of -5.21\% (or USD -1.2 billion), and even more so when these are data breaches. Instead, we cannot show any time-varying effect of cyber incidents or a specific effect of the type of news as had previously been advocated.\\ \\



\end{abstract}



\thispagestyle{empty}
\clearpage

\onehalfspacing
\setcounter{footnote}{0}
\renewcommand{\thefootnote}{\arabic{footnote}}
\setcounter{page}{1}

\section{Introduction}

\input{sections/introduction}

\section{Literature review}
\input{sections/literature_review}

\section{Data and methodology}
\input{sections/data_method}

\section{Results}
\input{sections/results}

\section{Discussion}
\input{sections/discussion}

\section{Conclusion}
\input{sections/conclusion}

\clearpage

\bibliographystyle{unsrtnat}
\bibliography{bibliography.bib}


\clearpage
\section*{Appendix}

\setcounter{table}{0}
\renewcommand{\thetable}{A\arabic{table}}
\setcounter{figure}{0}
\renewcommand{\thefigure}{A\arabic{figure}}

\input{sections/appendix}

\end{document}

%% file: sections/introduction.tex
Following the hack of ICBC, Marcus Murray, the founder of Swedish cybersecurity firm Truesec, declared:

\quotes{This is a true shock to large banks around the world. The ICBC hack will make large banks around the globe race to improve their defences, starting today.}\footnote{\url{https://time.com/6333716/china-icbc-bank-hack-usb-stick-trading/}}

This major event, which has cost millions to the firm itself and the clearing organization of the US treasury market, is a single example of how detrimental cyber incidents can be for the economy. Thus, quantifying these costs is an essential starting point for cybersecurity investments of target organizations and, on the other hand, for cyber insurers and cybersecurity service providers to determine their market size. Given the publicly available information that stock markets provide the researchers with, a popular econometric approach, the event study, an econometric method to estimate the counterfactual would an event not occur and compare with the actual realization, has been used to assess these costs systematically and explain their sources of variation. Since Campbell, Gordon, Loeb, et al., the first to adopt this approach for cyberattacks, a large strand of literature in the economics of information security has used it, with conflicting results at a granular level but yielding overall support for the view that cyberattacks or more broadly cyber incidents do affect firms' value \cite{CampbellGordonLoebZhou2003}.

In this paper, we revisit the impact of cyber incidents on listed firms using observed adjusted returns on their stocks and advanced event study methods. Although such studies already exist in cybersecurity and financial literature, we are the first to adopt the following setting. First, we use newswire headlines that we filter to ensure they bring relevant information regarding cyber incidents. Next, we use state-of-the-art estimation methods to account for the cross-correlation of errors at the estimation stage of our cumulative abnormal returns (CARs) using a seemingly unrelated regression (SUR) estimation. This ensures that the reported CAR coefficients are unbiased regarding economic magnitude. We find overall negative abnormal returns of about -0.89\% irrespective of the target firm or the type of cyber incident. Next, we adjust the standard errors using the Boehmer, Musumecci, and Poulsen and the Kolari and Pynnonen corrections for event-induced variance and cross-correlation, respectively \cite{BoehmerMusumecciPoulsen1991,KolariPynnonen2010}. This is essential and a significant caveat in the existing literature. First, the proximity of firms suffering from cyber incidents is strong regarding characteristics and behavior, thereby generating implicit cross-correlation in their returns. Second, it is likely that a cyber incident does not only affect the target firm but the entire market. These two biases would then favor an underestimation of standard errors.

We find that unadjusted standard errors yield CARs with significance well below the 1\% level. These results translate into overall economic costs of USD 277 billion market capitalization reduction over the period (USD 25 billion per year), with a median cost of USD 123 million for the US-listed firms at each cyber incident. 
However, once we account for the above-mentioned biases, we find no statistical significance remains at the aggregated level. This goes against the existing consensus in the event study literature on cyberattacks. We subsequently turn to a more granular analysis of these effects, adding variables controlling for the type of cyber incidents, the sector of the target firm, year-fixed effects, firm characteristics, and news sources.
We cannot identify any time pattern in the magnitude and statistical significance of abnormal returns, as found by Gordon, Loeb, and Zhou, in an earlier period \cite{GordonLoebZhou2011}. Moreover, some specific years have positive (albeit not statistically significant) abnormal returns.
In contrast, we identify a significant marginal effect on firms in the health sector, translating into an average loss of USD 1.2 billion per event and for cyber incidents involving data breaches (average loss of USD 1.9 billion). In particular, the interaction coefficient, which measures the combined effect of the firms belonging to the intersection of the two sub-samples, reaches as low as -7.07\% and is statistically significant at the 10\% level. In contrast, we cannot find evidence that ransomware attacks significantly affect firms' value, nor that the financial sector is more sensitive than others, among other non-statistically significant results.
We next test whether intrinsic firm characteristics, typical performance determinants in the financial and accounting literature, can explain the variations in abnormal returns. Without concluding, we test the firm size, age, book-to-market ratios, and price-to-earnings ratios together. In only one specification, the firm size and the price-to-earnings ratio weakly explain abnormal returns, with firm size (price-to-earnings ratio) being a positive (negative) determinant, but with statistical significance levels well above the 1\% level, a weak overall explanatory power (7.5\% at most), and a large share of variance captured by the intercept.
Finally, we test whether the source of the news has an impact on the abnormal returns. We control whether Reuters, Twitter, or another source first releases the news. Once again, the explanatory power does not allow us to reject the null of a particular effect.

Our contribution is threefold. First, we use a novel dataset of news headlines of cyber incidents between 2012 and 2022 filtered from the large Refinitiv news dataset, covering a more recent period than the existing literature, to estimate abnormal returns around cyber incidents. Second, we adopt the most advanced econometric estimation methods, such as the Seemingly Unrelated Regression, to avoid bias in the coefficients and standard errors adjusted for event-induced variance and cross-correlation to avoid statistical significance biases. Whereas our approach yields coefficients economically on par with previous event studies on cyberattacks, we cannot find any statistical significance at the usual levels, casting doubt on previous claims. Last, these results encourage us to dig further to identify possible significant determinants of our estimated abnormal returns. We test, in turn, the type of cyber incidents, the firm's sector, firm characteristics, the incident's year, and the news source. Only a few of these potential explanatory variables explain the variance of the CARs. However, we find evidence that data breaches are the sole type of incidents that firms should worry about, specifically if they belong to the healthcare sector. We argue that cyber incidents are not as detrimental to firms' value overall as previously advocated, except for some specific situations.

The remainder of the paper proceeds as follows. Section 2 introduces the existing literature and develops related hypotheses. Section 3 presents the data and methods, Section 4 details the results, and Section 5 concludes.

%% file: sections/literature_review.tex
\subsection{Event study methodology and critical results}

\subsubsection{Methodology}

Event study methods are heavily employed in all branches of economics. The first modern form of the empirical test to assess the impact of one or several economic events on one or several data series dates back to Ball and Brown, whose research focuses on the impact assessment of accounting measures changes on stock prices \cite{BallBrown1968}.\footnote{Dolley is alternately cited as the first author to employ event studies on stock splits \cite{Dolley1933}.}
Fama, Fisher, Jensen, et al. also help define the modern empirical approach to event studies, as they study the abnormal returns around stock split announcements to falsify the efficient market hypothesis \cite{FamaFisherJensenRoll1969}. Although they find significant positive abnormal returns around these announcements, with a market model, they also rationalize them.

While studying the effects of the accuracy of 336 forecast annual earnings disclosures on stock prices over the 1963--1967 period, Patell introduces a first adjustment in the event study methods \cite{Patell1976}. He proceeds by scaling the standard errors used for the t-test of significance by the standard deviation of the estimation period residuals. Boehmer, Musumecci, and Poulsen propose a similar adjustment to scale standard errors based on the event-induced variance (see also, Savickas) \cite{BoehmerMusumecciPoulsen1991,Savickas2003}. The aforementioned adjustments are the first to account for the event-induced heteroskedasticity of the time series under consideration.

On the other hand, empirical adjustments may be needed to account for the cross-sectional dependence of several firms affected by the same event. To alleviate this effect, Malatesta proposes the joint generalized least squares approach \cite{Malatesta1986}. Using simulations, he cannot demonstrate, however, that this approach dominates more simplistic cross-sectional dependence treatment. Similarly, Salinger takes the standpoint that actual abnormal returns must be uncorrelated even though their estimates are not \cite{Salinger1992}. He derives an adjustment formula for the standard errors using a standard market model approach and a dummy variable. Applying this correction to an event study of post-merger performance, he shows that omitting the adjustment can lead to severe over-rejection of the null hypothesis of no abnormal returns. Similarly, Karafiath develops the dummy variable approach, which consists of appending a vector of zeros (ones) outside (inside) the event window \cite{Karafiath1994}. This approach enables obtaining abnormal returns in a single step, allowing for more straightforward adjustments of the standard errors and an easier interpretation of the residuals.

Another drawback of event studies is that they often rely on daily returns, which are more prone to autocorrelation, leading to underestimation of standard errors. Brown and Warner propose a series of adjustments to address this issue and discuss the cross-sectional dependence and event-induced variance issues \cite{BrownWarner1985}.

The studies above and the state-of-the-art methods generally apply to abnormal returns of short windows around the event considered (CARs). However, another strand of literature recognizes issues from estimating longer windows around and after the events, the so-called buy-and-hold abnormal returns (BHAR). Advances in treating such BHAR are discussed in Campbell, Lo, and MacKinley and Kothari and Warner \cite{CampbellLoMacKinley1997,KothariWarner2007}.

Finally, Kolari and Pynnonen develop a t-statistics that summarizes and treats all the aforementioned CAR adjustments proposed \cite{KolariPynnonen2010}. It deals jointly with event-induced variance, cross-correlation, and autocorrelation biases. They test the unbiasedness of their standard errors and the power of the tests in simulations and generate a test that is the most effective available parametric test to date.\footnote{This literature review voluntarily skips the minor literature on non-parametric tests. See, \textit{e.g.}, the non-parametric rank test of Kolari and Pynnonen\cite{KolariPynnonen2011}.}\footnote{Also see, Lee and Varela, for tests of the different available standard errors adjustments \cite{LeeVarela1997}.}.

\subsection{Cyber incident costs}

\subsubsection{Abnormal returns around cyber incidents}

Gordon, Loeb, and Zhou assess the impact of information security breaches on stock returns by estimating CARs over a three-day event window centered on news of cybersecurity incidents \cite{GordonLoebZhou2011}. They find that news about information security breaches generates significant negative CARs for publicly traded firms. They additionally uncover a significant downward shift in the impact of security breaches in the post-9/11 sub-period. They interpret these findings with a shift in investors' attitudes towards cyber breaches. In a similar study, Campbell, Gordon, Loeb, et al. uncover highly significant (non-significant) negative CARs for information security breaches (not) involving unauthorized access to confidential data \cite{CampbellGordonLoebZhou2003}.

The aforementioned studies estimating the effects of cyber incidents on CARs do not adjust their results for the presence of cross-correlation (SUR estimations of Zellner) and do not adjust the standard errors for event-induced variance (Boehmer, Musumecci, and Poulsen) and cross-correlation (Kolari and Pynnonen) \cite{Zellner1962,BoehmerMusumecciPoulsen1991,KolariPynnonen2010}. They also find support for time-varying effects that we can test in a more recent subsample. Thus, we formulate our first null hypotheses as follows:

\begin{itemize}
    \item $H_{1a}$: Cyber incidents do not generate economically significant CARs either with OLS estimation or with adjustments of event-study dummy coefficients for cross-correlation (SUR estimation).

    \item $H_2$: The CARs' statistical significance is insensitive to adjustment that considers cross-correlation and event-induced variance.

    \item $H_{3}$: Abnormal returns around cyber incidents are time-invariant.
\end{itemize}

Johnson, Kang, and Lawson also show that, on average, publicly traded firms in the U.S. face CARs of about -0.37\% over a data breach \cite{JohnsonKangLawson2017}. Additionally, breaches resulting from payment card fraud contribute more to negative abnormal returns than other breach types. Lastly, they find a positive link between the magnitude of these CARs and the card breach's size.
Lending, Minnick, and Schorno relate corporate governance and social responsibility to the probability of data breaches \cite{LendingMinnickSchorno2018}. Measuring changes in stock returns after breaches, they uncover a persistent effect of –3.5\% of one-year BHAR. They also find that banks with breaches significantly decrease deposits, while non-banks experience large sales decreases. Tosun studies how financial markets react to unexpected corporate security breaches in both the short- and the long-run \cite{Tosun2021}. He uncovers that market participants anticipate negative stock price changes while analyzing selling pressure and liquidity measures. However, the negative effect is significantly negative only the day following the security breach public announcements and is linked to an adverse impact on the firm reputation. Kamiya, Kang, Kim, et al. find additional support for a reputation loss of targeted firms through a drop in credit ratings or a decrease in sales growth \cite{KamiyaKangKimMilidonisStulz2021}. Based on these results, we argue that our novel and updated dataset, along with the most advanced methods, calls for testing the following null hypothesis and sub-hypotheses,

\begin{itemize}
    \item $H_{4}$: CARs around cyber incidents cannot be explained by:
    \begin{itemize}
        \item $H_{4a}$: Type of cyber incidents (ransomware, data breach, security breach, etc.)
        \item $H_{4b}$: Type of sector the target firm belongs to (technology, health, financial, etc.)
        \item $H_{4c}$: Typical firm characteristics such as size, book-to-market ratio, price-earnings ratio, and firm age
    \end{itemize}
\end{itemize}

Andreadis, Kalotychou, Louca, et al. investigate how information dissemination about cyberattacks through major news sources affects municipalities' access to finance, particularly in the municipal bond market \cite{AndreadisKalotychouLoucaLundbladMakridis2023}. Using a differences-in-differences framework, they show that an increase in the number of cyberattacks covered by news articles at the county level and the corresponding number of cyberattack news articles significantly adversely impact municipal bond yields. A 1\% rise in the number of cyberattacks covered in news articles results in offering yields increasing by 3.7 to 5.9 basis points, depending on the level of cyberattack exposure. In our case, testing for the number of news articles as a determinant of CARs would imply the use of forward-looking information and thus bias our results. However, with our sample of the first available news, we can test for the impact of the news source. Thus, we define our last null hypothesis as follows:

\begin{itemize}
    \item $H_5$: The news source does not drive the impact of cyber incidents on CARs.
\end{itemize}

Finally, other determinants of abnormal returns around cyber incidents have been used, for which our dataset does not allow testing. They include Jensen and Paine, who use data on municipal IT investments, ransomware attacks, and bond performance \cite{JensenPaine2023}. They cannot find an immediate impact on bond yields within 30 days of a cyberattack. However, over the subsequent 24 months post-cyberattack, municipal bond yields gradually decreased while IT spending increased. They argue that this decline in bond yields is driven by a reduction in cyber risk due to increased IT investment.
Gordon, Loeb, and Sohail examine how voluntary information security disclosures impact firm value, using a dataset of 1,641 firms that disclose such information and 19,266 that do not \cite{GordonLoebSohail2010}. They find that these disclosures can reduce litigation costs and lower a firm's cost of capital by reducing information asymmetry between management and investors. They find a positive relationship between voluntary information security disclosures and firm value. Firms that disclose this information also exhibit narrower bid-ask spreads than those that do not. Hilary, Segal, and Zhang also study cyber risk disclosures and uncover that the market reaction to cyber breaches is statistically significant but economically narrow \cite{HilarySegalZhang2016W}.

\subsubsection{Other methods for cyber incident costs estimation}

Bouveret studies the global cyber risk for the financial sector and the various types of cyber incidents \cite{Bouveret2018W}. Using a Value at Risk (VaR) framework, he uncovers an average country loss from cyberattacks of USD 97 billion and a VaR between USD 147 and 201 billion. He argues that essential potential aggregated losses exist in the financial sector, several orders of magnitude higher than the cyber insurance market could cover. Romanosky delves into the nature and expenses related to cyber events \cite{Romanosky2016}. His analysis of a dataset comprising over 12,000 incidents reveals a skewed cost distribution, with an average cost of USD 6 million and a median cost of USD 170,000, akin to a typical annual IT security budget for firms. This leads to the hypothesis that firms may be operating optimally secure due to the relatively low costs, thus investing modestly in data protection measures. Other related studies include \textit{e.g.}, Anderson, Barton, Böhme, et al. (2013) and Anderson, Böhme, Clayton, et al. (2019) \cite{AndersonBartonBöhmeClaytonEetenLeviMooreSavage2013,AndersonBartonBöhmeClaytonGananGrassoLeviMooreVasek2019}.

A last strand of research in finance and cybersecurity uses asset pricing techniques and the cross-sectional analysis of listed stocks. Even though this literature relates more to the perceived risk than actual realization, this literature is worth mentioning and includes \textit{e.g.}, Florackis, Louca, Michaely, et al., Jamilov, Rey, and Tahoun, Jiang, Khanna, Yang, et al., and Liu, Marsh, and Xiao who all find that the cyber risk is priced to some extent in the cross-section of stock returns \cite{FlorackisLoucaMichaelyWeber2023,JamilovReyTahoun2021W,JiangKhannaYangZhou2023,LiuMarshXiao2022W}.\newline\\

%% file: sections/data_method.tex
\subsection{Market data}
\label{section:md}

We download public equity data from Wharton Research Data Services (WRDS), using the Center for Research in Security Prices (CRSP) and S\&P Global Market Intelligence's Compustat database. We report the list of variables in Table \ref{tab:variable_descriptions}. We write a Python script that queries all available information from WRDS' API and filters the firms so that all of the retained firms have at least one cyber incident relating to them in our news dataset described below. We extract daily stock returns and financial ratios for 119 firms between January 2012 and December 2022. We also download the one-month Treasury bill rate and returns on the market, book-to-market (HML), and size (SMB) factors from the Kenneth French data repository\footnote{Available at: \href{http://mba.tuck.dartmouth.edu/pages/faculty/ken.french/data\_library.html}{http://mba.tuck.dartmouth.edu/pages/faculty/ken.french/data\_library.html}}.

\subsection{News data}
\label{section:methodology}

\subsubsection{News treatment}

\begin{figure}[H]
    \noindent\makebox[\textwidth]{%
    \includegraphics[width=0.75\textwidth]{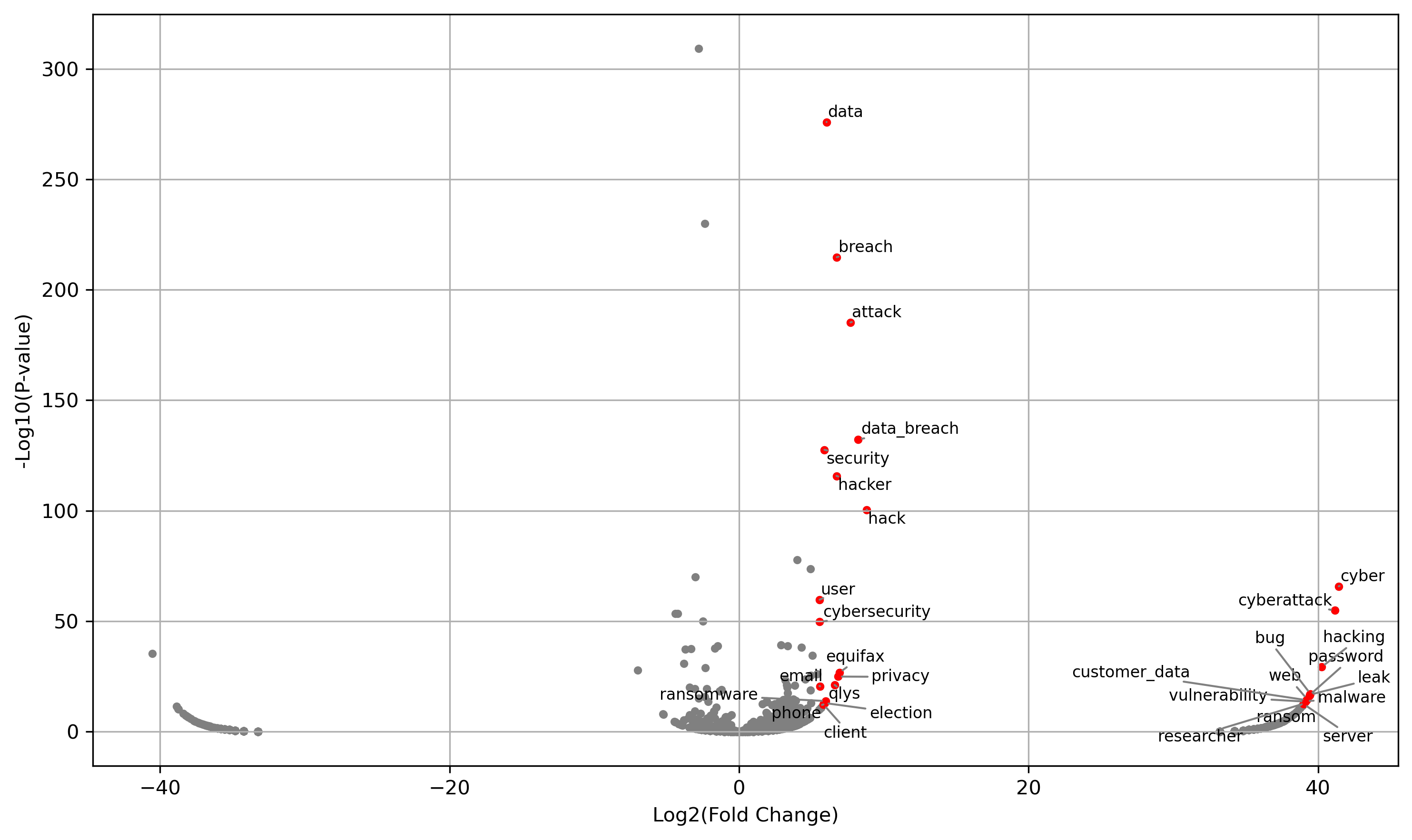}
    }
     \caption{\textbf{Volcano plot of cyber incident headlines against normal headlines}}\bigskip
     \footnotesize{Volcano plot of a normal news corpus against a cybercrime news corpus. The population of both samples is 12,297 and is extracted from Refinitiv. The x-axis represents a log2 fold change, while the y-axis depicts the negative logarithm (base 10) of the p-value derived from a chi-squared test. The x and y thresholds are set to optimally select the relevant keywords.}
     \label{fig:volcano}
\end{figure}

We collect news headlines from the Refinitiv Eikon platform, which is recognized for its comprehensive financial news coverage and use in previous research. This platform categorizes news with a \quotes{Cybercrime} tag, focusing on cyber incidents. We begin with the download of 106,248 headlines. We first filter this dataset by keeping only English headlines related to listed companies, reducing our sample to 27,244. Second, we keep only North American firms using their CUSIPs and narrow the sample to 12,561. To identify characteristic keywords in cyber incident headlines, we compare 12,297 of these headlines against a normal news corpus using a volcano plot, as illustrated in Figure \ref{fig:volcano}.

The volcano plot contrasts word frequencies between two distinct datasets. The x-axis shows the log2 fold change, indicating the relative frequency shift of words between cybercrime and regular headlines. Words more common in cybercrime headlines have a higher positive value. The y-axis presents the negative log (base 10) of the p-value from a chi-squared test, highlighting the statistical significance of the frequency differences. Words at the top of the plot show substantial frequency changes, thus distinguishing cybercrime headlines from regular news.

Based on this analysis, we identify keywords typically associated with cyber incidents, such as \textit{breach, hacker, cybersecurity, ransomware, malware, leak, vulnerability, data, and attack}. These terms, though not exhaustive, cover a broad range of cyber incident scenarios (\textit{e.g.}, \textit{[Company Name] is experiencing a [Keyword]}). Applying these keywords, we filter the headlines to include specific cyber terms and company names, making exceptions for Meta and Alphabet to capture incidents related to Google or Facebook. This process yields 3,606 headlines.

We then use the MoritzLaurer/DeBERTa-v3-large-mnli-fever-anli-ling-wanli model from the Hugging Face platform for text entailment.\footnote{Available at \href{https://huggingface.co/MoritzLaurer/DeBERTa-v3-large-mnli-fever-anli-ling-wanli}{https://huggingface.co/MoritzLaurer/DeBERTa-v3-large-mnli-fever-anli-ling-wanli}} This is to validate headlines whose content is more complex. For instance, \textit{[Company Name] has been under a cyber incident or attack.} We seek an entailment score above 93\% to underfit the results for later manual evaluation. Based on BERT and DeBERTa, this model is fine-tuned on various natural language inference datasets, offering 885,242 hypothesis-premise pairs, and is highly rated on Hugging Face as of June 6, 2022. This step results in 1,465 headlines.

\begin{figure}[H]
    \noindent\makebox[\textwidth]{%
    \includegraphics[width=\textwidth]{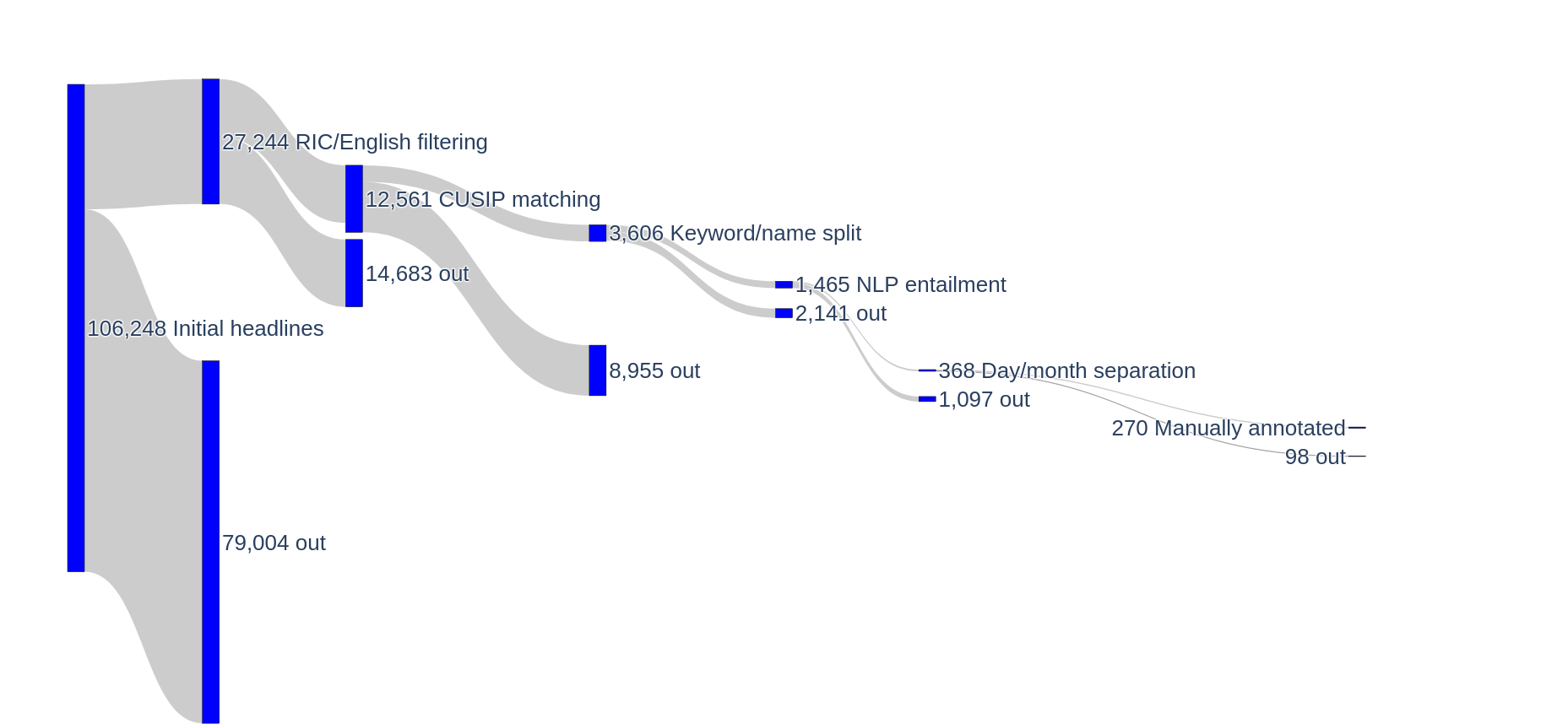}
    }
     \caption{\textbf{Filtering process of cyber incident headlines}}\bigskip
     \footnotesize{This diagram depicts the filtering steps of the original cyber incident headline set from Refinitiv. We only keep headlines in English that mention listed companies with an identifier (RIC). Next, we keep North American listed companies that have a CUSIP. We filter based on cyber incident keywords. We subsequently check whether the headline is a premise to the following entailment question: \quotes{[Company Name] has been under a cyber incident or a cyberattack.} We additionally keep the earliest headline per day and company. These must be two months apart for a given company (starting with the earliest). Finally, we manually change the dates and drop headlines that are not about a cyber incident or are not the first to mention it.}
     \label{fig:sankey_preprocessing}
\end{figure}

To avoid duplication, we retain only each company's first headline per day. These headlines must lie two months apart for the same company, with exceptions for corporations like Meta, Alphabet, Google, and Apple due to their higher frequency of incidents. This results in 368 headlines. Finally, we manually verify these headlines for date accuracy, relevance, and attack type, resulting in a final count of 270 headlines, for which we obtain the corresponding information regarding the target company name and identifiers, type of cyber incident, date, and name of the news source. We depict all the aforementioned filtering steps in Figure \ref{fig:sankey_preprocessing}.

\begin{figure}[H]
    \noindent\makebox[\textwidth]{%
    \includegraphics[width=\textwidth]{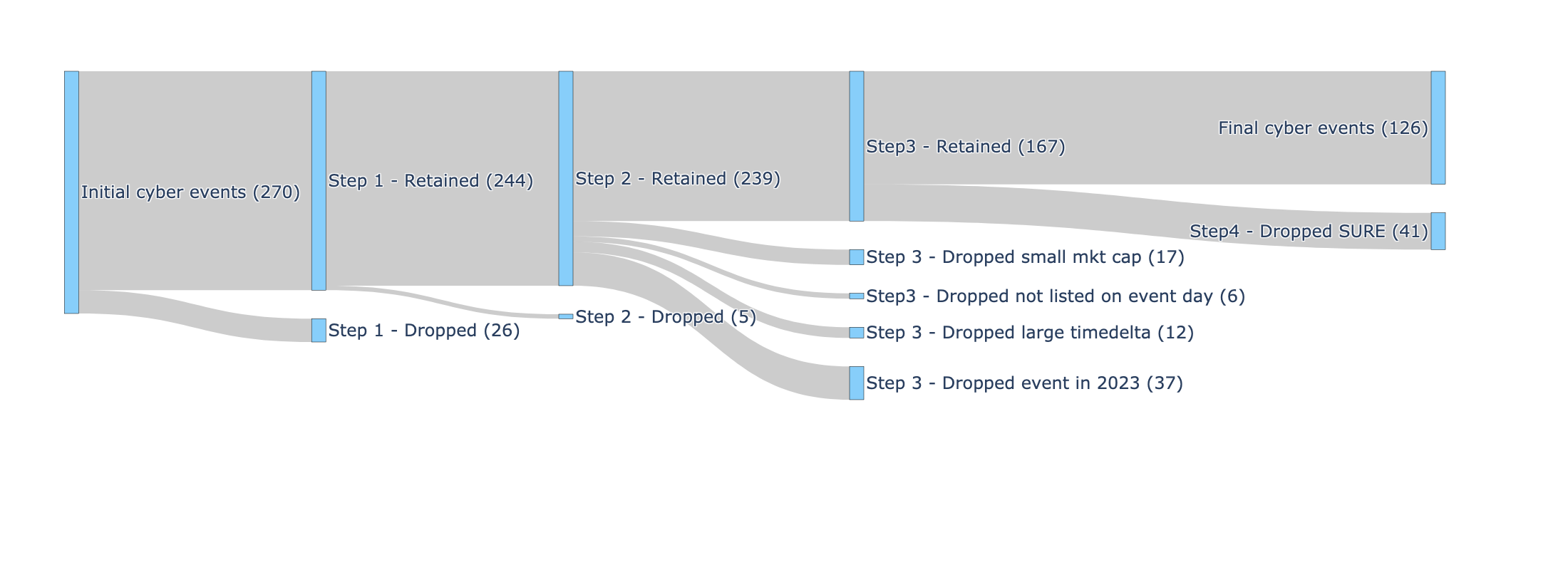}
    }
     \caption{\textbf{Evolution of the number of incidents}}\bigskip
     \footnotesize{In step 1, we drop all firms not in the WRDS database. In step 2, we drop all incidents affecting cybersecurity-providing firms. In step 3, we drop incidents where the firm's market capitalization is lower than \$300 million, the firm is not listed during the incident, the incident is in 2023, or the incident window is larger than five days. In step 4, we drop all firms not listed between 2013.12.19 and 2022.10.13 and/or whose cyber incident(s) did not occur between those dates.}
     \label{fig:sankey_events}
\end{figure}

\begin{figure}[H]
    \noindent\makebox[\textwidth]{%
    \includegraphics[width=\textwidth]{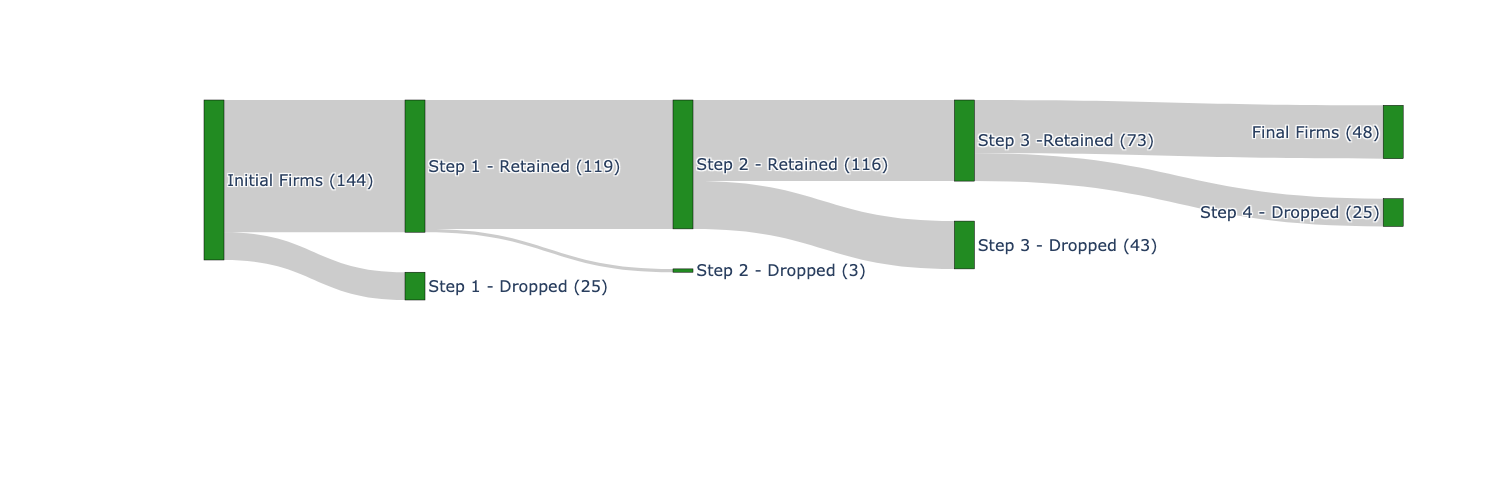}
    }
     \caption{\textbf{Evolution of the number of firms}}\bigskip
     \footnotesize{In step 1, we drop all firms not in the WRDS database. In step 2, we drop all incidents affecting cybersecurity-providing firms. In step 3, we drop incidents where the firm's market capitalization is lower than \$300 million, the firm is not listed during the incident, the incident is in 2023, or the incident window is larger than five days. In step 4, we drop all firms not listed between 2013.12.19 and 2022.10.13 and/or whose cyber incident(s) did not occur between those dates.}
     \label{fig:sankey_firms}
\end{figure}

 We further filter the number of events by merging them with the WRDS and Compustat databases. Hence, we drop all firm-related events not in the WRDS database. We also drop firms with very small market capitalizations (below \$300 million), firms not listed on the event day, and firms where the event happens right before or after a long weekend. We end up with 167 events relating to 73 firms. These filtering steps are illustrated in Figures \ref{fig:sankey_events} and \ref{fig:sankey_firms}.

\begin{figure}[H]
    \noindent\makebox[1.09\textwidth]{%
    \includegraphics[width=0.5\textwidth]{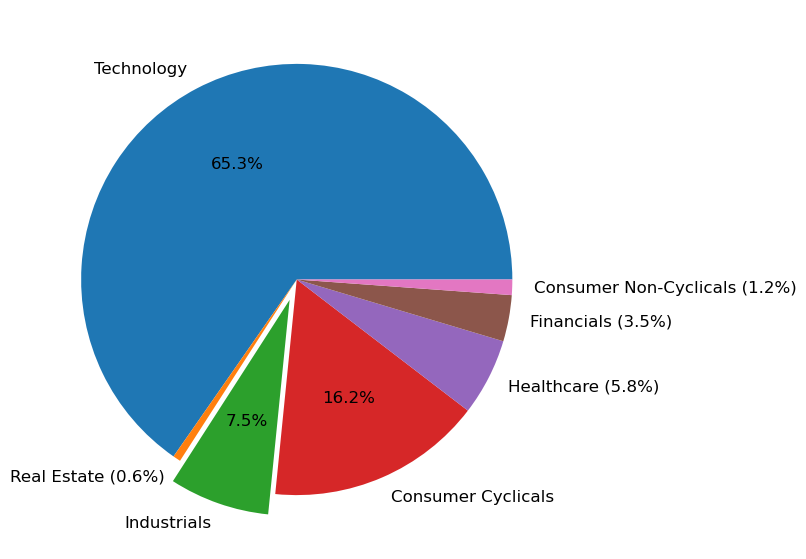}
    }
     \caption{\textbf{Sector distribution}}\bigskip
     \footnotesize{Sector distribution of the firms remaining after step 3 on Figure \ref{fig:sankey_firms}. We use Refinitiv's sector classification.}
     \label{fig:indsutry_dist}
\end{figure}

\begin{figure}[H]
    \noindent\makebox[\textwidth]{%
    \includegraphics[width=0.5\textwidth]{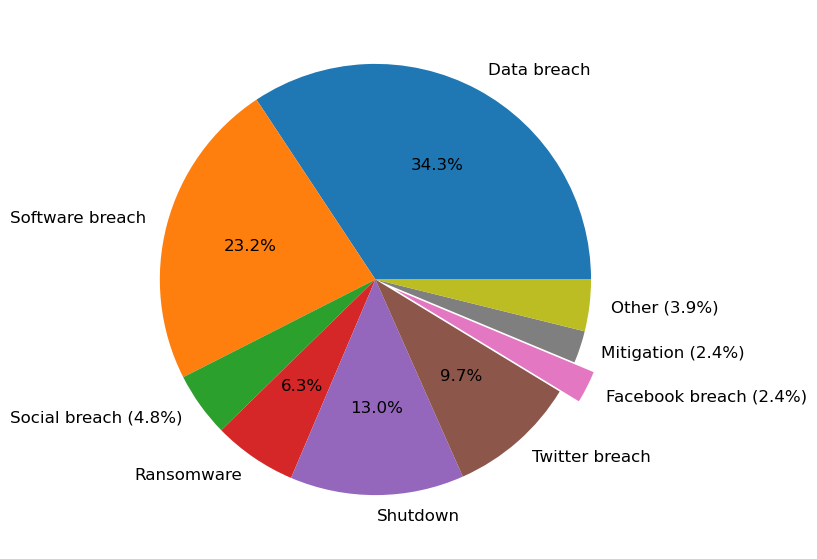}
    }
     \caption{\textbf{Attack distribution}}\bigskip
     \footnotesize{Distribution of the different types of attacks of the incidents remaining after step 3 on Figure \ref{fig:sankey_events}. The \quotes{Other} category regroups the Cyber breach (1.93\%) and Stolen funds (1.93\%) categories.}
     \label{fig:attack_dist}
\end{figure}

We depict the sector and attack type distributions in Figures \ref{fig:indsutry_dist} and \ref{fig:attack_dist}, respectively. Most firms are from the technology sector, whereas the types of attacks are more diversified. The dataset's two most common types of attacks are data breaches and software breaches.

\subsubsection{Abnormal returns}

We compute abnormal returns with the Fama and French three-factor model (FF3) using Eq. \ref{eq:AR_regression} below \cite{FamaFrench1992}.

\begin{equation}
    \begin{split}
     R_i^e & = \alpha_i + \beta_i^{Mkt} R_{Mkt}^e + \beta_i^{SMB} R_{SMB} + \beta_i^{HML} R_{HML} + \sum\limits_{j = 0}^{k} AR_{event_j}\mathds{1}_{event_j}\\
    \end{split}
    \label{eq:AR_regression}
\end{equation}

where $R_{Mkt}^e$ is the excess return on the market portfolio, $R_{SMB}$ is the return on the size factor, $R_{HML}$ is the return on the book-to-market factor, $k$ is the total number of events in the dataset, $AR_{event_j}$ is the abnormal return on event $j$ and $\mathds{1}_{event_j}$ is a dummy vector that takes the value \quotes{1} on the day of event $j$ and \quotes{0} otherwise. This equation can be written as $R_i^e = X \beta_i$ in matrix notation. We estimate Eq. \ref{eq:AR_regression} for each firm, using Ordinary Least Squares (OLS) regressions and Seemingly Unrelated Regressions (SUR) (see Zellner; 1962) \cite{Zellner1962}. In the SUR estimation, the error terms are assumed to be correlated across the equations, and each equation must have the same number of observations. Hence, we perform SUR on a subset of the data with the most cyber incidents relating to firms with daily returns for all trading days in the subset. We find the optimal subset to start on the 19th of December 2013 and end on the 13th of October 2022. There are 126 cyber incidents relating to 48 firms in this subset of the data, as can be seen in Figures \ref{fig:sankey_events} and \ref{fig:sankey_firms}. We also compute abnormal returns with the zero-benchmark model presented in Eq. \ref{eq:AR_regression_zb}.

\begin{equation}
    \begin{split}
     R_i^e & = \alpha_i + \sum\limits_{j = 0}^{k} AR_{event_j}\mathds{1}_{event_j}\\
    \end{split}
    \label{eq:AR_regression_zb}
\end{equation}

We compute CARs over a three-day window centered on cyber incidents.

\subsubsection{t-stat adjustments}

We use scaled abnormal returns and two t-test statistics from Kolari and Pynnonen (2010) \cite{KolariPynnonen2010}. We borrow their notations for the remainder of the paper. The scaled abnormal returns can be computed using Eq. \ref{eq:scaled_AR},

\begin{equation}
    \begin{split}
     A_{i,t} & = \frac{AR_{i,t}}{s_i\sqrt{1+x_t'(X'X)^{-1}x_t}}\\
    \end{split}
    \label{eq:scaled_AR}
\end{equation}

where $s_i$ is the regression residual standard deviation, X is the matrix of explanatory variables from the matrix notation of Eq. \ref{eq:AR_regression} and $x_t$ the $t$-th row of X. A feasible estimator of the variance of the scaled abnormal returns can be computed using Eq. \ref{eq:SAR_variance},

\begin{equation}
    \begin{split}
     s_A^2 & = \frac{s^2}{1-\overline{r}},\\
    \end{split}
    \label{eq:SAR_variance}
\end{equation}
where $s^2$ is the sample cross-sectional variance of event-day scaled abnormal returns and  $\overline{r}$ is the average sample cross-correlation of the residuals.

The ADJ-BMP statistic is computed as shown in Eq. \ref{eq:ADJ_BMP}, below.

\begin{equation}
    \begin{split}
     t_{AB} =  \frac{\bar{A} \sqrt{n}}{s_A \sqrt{1+(n-1)\bar{r}}},\\
    \end{split}
    \label{eq:ADJ_BMP}
\end{equation}
where $n$ is the number of firms in the sample. This statistic is residual cross-correlation and event-induced volatility-adjusted.

The ADJ-PATELL statistic is computed as shown in Eq. \ref{eq:ADJ_PATELL}.

\begin{equation}
    \begin{split}
     t_{AP} =  \frac{\bar{A} \sqrt{n}}{\sqrt{(m-p-1)/(m-p-3)}\sqrt{1+(n-1)\bar{r}}},\\
    \end{split}
    \label{eq:ADJ_PATELL}
\end{equation}
where $p$ is the number of explanatory variables in the expected return regression (three in our case) and $m$ is the number of days in the sample. This statistic is residual cross-correlation-adjusted.

%% file: sections/results.tex
\subsection{CAR estimations}

\begin{figure}[H]
    \noindent\makebox[1.1\textwidth]{%
    \includegraphics[width=0.55\textwidth]{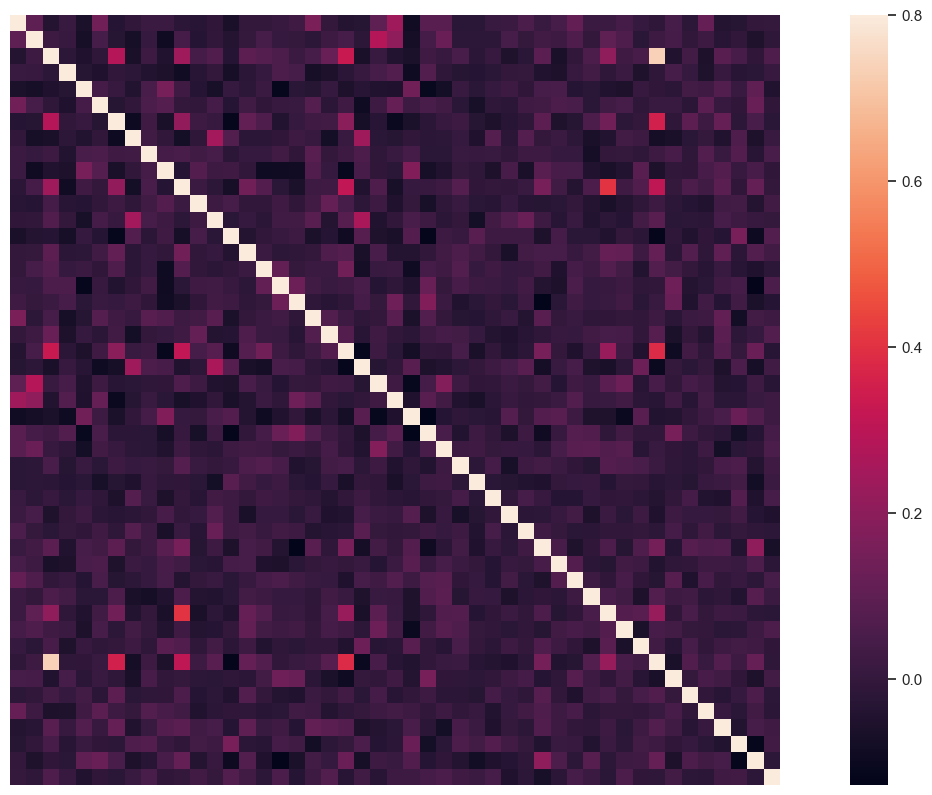}
    }
     \caption{\textbf{Correlation of residuals}}\bigskip
     \footnotesize{Correlation matrix of residuals, obtained by estimating Eq. \ref{eq:AR_regression}.}
     \label{fig:residual_corr}
\end{figure}

Figure \ref{fig:residual_corr} presents the correlation matrix of the residuals obtained from estimating Eq. \ref{eq:AR_regression}. Overall, the residuals do not have high correlations; the average correlation is 0.012. Since the SUR estimation is equivalent to OLS when the error terms are uncorrelated between equations, we expect the results obtained with the SUR estimation to be similar to the ones obtained with OLS.

Table \ref{tab:average_car_tstat} presents the average CARs and unadjusted and adjusted t-statistics. 

\begin{table}[H]
    \begin{adjustbox}{width= \textwidth,center}
    \begin{tabular}{p{6cm}cccccc}
        & \multicolumn{3}{c}{Zero-benchmark} & \multicolumn{3}{c}{FF3 benchmark} \\
        \cmidrule(lr){2-4}
        \cmidrule(lr){5-7}
        & OLS & OLS limited & SUR & OLS & OLS limited & SUR  \\
        \midrule 
        $\overline{CAR}$ & -0.89\% & -0.88\% & -0.79\% & -0.74\% &  -0.84\% & -0.78\% \\
        \\
        Unadjusted t-stat & -1.949 & -2.258 & -2.325 & -1.856 & -2.579 & -2.434 \\
        ADJ-PATELL & - & -0.704 & -0.455 & - & -2.095 & -1.807\\
        ADJ-BMP & -0.157 & -0.323 & -0.273 & -0.698 & -1.226 & -1.033\\
        \midrule
        Number of incidents & 167 & 126 & 126 & 167 & 126 & 126 \\
        Number of days in regression & - & 2219 & 2219 & - & 2219 & 2219\\ 
        \bottomrule
    \end{tabular}
    \end{adjustbox}
    \caption{\textbf{Average CAR and t-stats}}\bigskip
    \footnotesize{$\overline{CAR}$ is the average cumulative abnormal return over a three-day window around the incidents, relative to the zero benchmark model (shown in Eq. \ref{eq:AR_regression_zb}) or the three-factor model of Fama and French (shown in Eq. \ref{eq:AR_regression}) \cite{FamaFrench1992}. ADJ-BMP and ADJ-PATELL refer to the adjusted statistics of the same name from Kolari and Pynnonen \cite{KolariPynnonen2010}. The ADJ-PATELL statistic is cross-correlation-adjusted, and the ADJ-BMP statistic is cross-correlation and volatility-adjusted. The OLS model allows for the number of days in the regression to differ for each firm, and the average number of days is 2398. The OLS limited model is the OLS model restricted to the same period, the same subset of firms, and the same incidents as the SUR model.}
    \label{tab:average_car_tstat}
\end{table}

We observe that the average CARs obtained with the SUR approach are close to the ones obtained with OLS. The average CAR for the three-factor benchmark is economically significant, at around -0.8\%, and statistically significant, at least at the 10\% level, when using unadjusted t-statistics. These results are close to the ones obtained in the existing literature over the past twenty years, and in particular, to the most recent event study on cyberattacks by Kamiya, Kang, Kim, et al., who find an average CAR of -0.768\% for the three-factor benchmark \cite{KamiyaKangKimMilidonisStulz2021}. Thus, we reject our null hypothesis $H_1$ regarding the economic significance of CARs. The t-statistics decrease when adjusting for residual cross-correlation and event-induced volatility. Importantly, none of the average CARs are statistically significant using the ADJ-BMP statistic, that is, once all biases are accounted for. Hence, the previously claimed statistical significance of CARs around cyberattacks seems not to hold once we properly adjust standard errors for the induced market-level effects of the cyber incidents. We are therefore not able to reject our hypothesis $H_2$.

\begin{figure}[H]
    \noindent\makebox[\textwidth]{%
    \includegraphics[width=\textwidth]{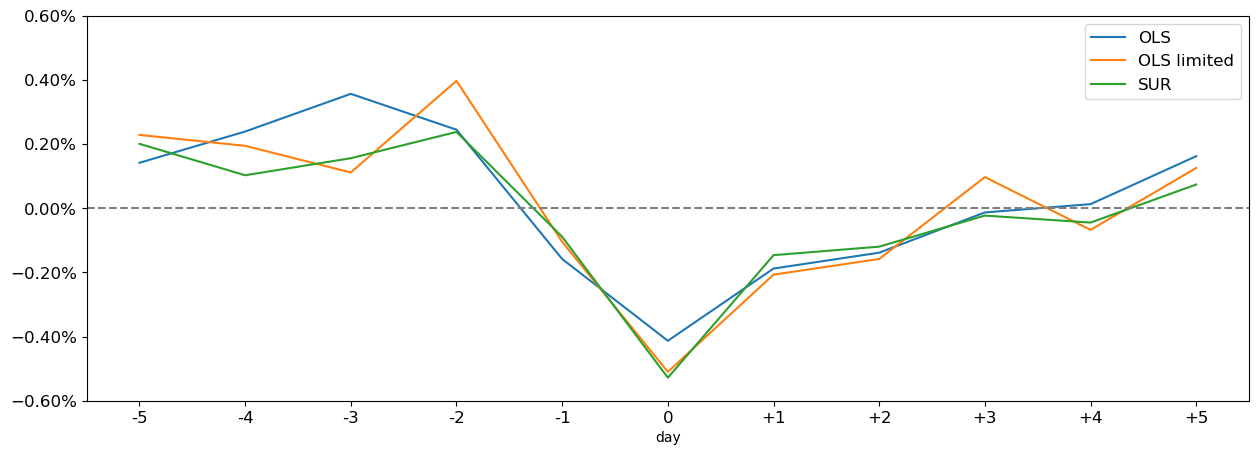}
    }
     \caption{\textbf{Average abnormal returns}}\bigskip
     \footnotesize{Average abnormal daily returns in an 11-day window centered on the cyber incidents. Day 0 is the day of the cyber incident. The OLS limited model is the OLS model restricted to the same period, the same subset of firms and incidents as the SUR model. Abnormal returns are computed with the three-factor model of Fama and French (1992), following Eq. \ref{eq:AR_regression} \cite{FamaFrench1992}.}
     \label{fig:AAR}
\end{figure}

\begin{figure}[H]
    \noindent\makebox[\textwidth]{%
    \includegraphics[width=\textwidth]{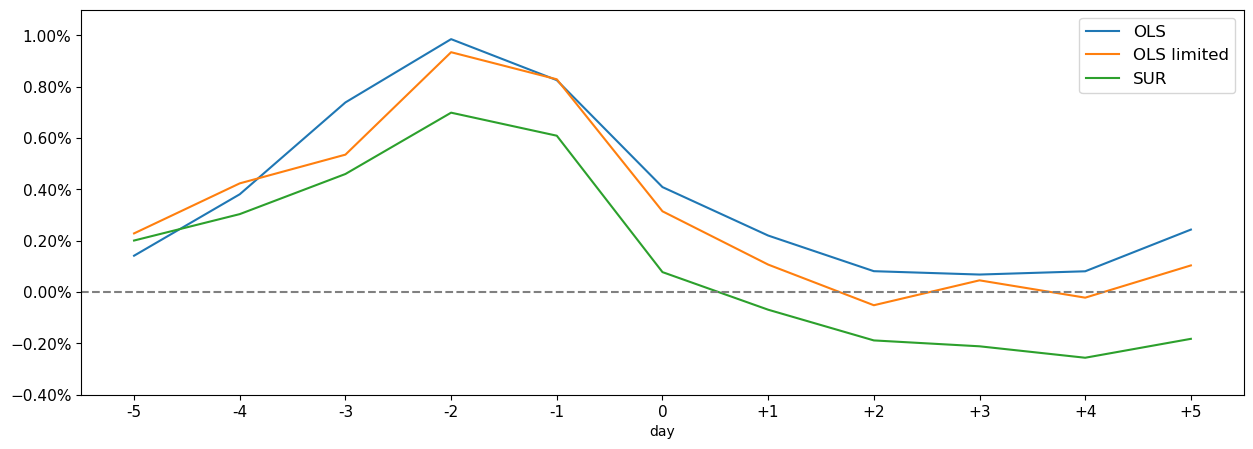}
    }
     \caption{\textbf{Cumulative average abnormal returns}}\bigskip
     \footnotesize{Cumulative average abnormal daily returns in an 11-day window centered on the cyber incidents. Day 0 is the day of the cyber incident. The OLS limited model is the OLS model restricted to the same period, the same subset of firms and incidents as the SUR model. Abnormal returns are computed with the three-factor model of Fama and French, following Eq. \ref{eq:AR_regression}. \cite{FamaFrench1992}}
     \label{fig:CAAR}
\end{figure}

We further investigate the evolution of abnormal returns around cyber incidents. Figure \ref{fig:AAR} shows the average abnormal return for each day in an 11-day window around cyber incidents. We observe that the abnormal returns drop up to three days after the cyber incident day but recover soon after. However, this is not a price recovery since post-event returns remain around zero. Figure \ref{fig:CAAR} shows that the CARs are at zero at the end of the window, but this is more due to a systematic increase before the event rather than a price recovery.\footnote{To verify that a misspecified benchmark does not drive our pre-incident positive returns, we have adopted several benchmarks in turn, including a zero benchmark that we present and discuss below. All of them show the same pattern. One plausible explanation is that technology firms are more prone to cyber incidents (or to report them) and that this subsample had tremendous growth over our sample period, which we observe pre-incident.}

Thus, it is difficult to conclude whether cyber incidents are informative regarding the firm's prospects or if they are purely temporary for technical (or behavioral) reasons, as advocated by Shleifer and Vishny \cite{ShleiferVishny1997}. While this is not the core focus of the study, this further contributes to our claim that the significance of cyber incidents may not have such a dramatic effect on the firm as previously found in some research.

\subsection{CARs through time}

To test our hypothesis $H_3$ regarding the time invariance of CARs around cyber incidents, we add year fixed-effect dummies in a panel regression of CARs. We use the CARs arising from the OLS and SUR estimations to ensure our results would not be driven by potential residual cross-correlation. We report our results in Table \ref{tab:CAR_on_years}.

\begin{table}[h]
    \begin{adjustbox}{width= 0.30\textwidth,center}
    \begin{tabular}{lcc}
        \hline
        \hline
        \multicolumn{3}{c}{Dependent variable: CAR}\\
        \hline
        & OLS & SUR \\
        \cmidrule(lr){2-2}
        \cmidrule(lr){3-3}
        & Model 3 & Model 3\\
        \hline
        2013 & \textbf{0.557} & -\\
        & [0.336]& -\\
        2014 & \textbf{-0.029} & -\\
        & [-0.017] & -\\
        2015 & \textbf{-1.109} & \textbf{-0.899}\\
        & [-0.598] & [-0.659]\\
        2016 & \textbf{-1.019} & \textbf{-1.760}$^{*}$\\
        & [-0.727] & [-1.756]\\
        2017 & \textbf{-0.324} & \textbf{-0.662}\\
        & [-0.262] & [-0.709]\\
        2018 & \textbf{-1.536} & \textbf{-1.790}$^{**}$\\
        & [-1.243] & [-1.981]\\
        2019 & \textbf{-0.651} & \textbf{-0.028}\\
        & [-0.430] & [-0.022]\\
        2020 & \textbf{-0.268} & \textbf{0.446}\\
        & [-0.260] & [0.550]\\
        2021 & \textbf{-1.799} & \textbf{-1.527}\\
        & [-1.456] & [-1.581]\\
        2022 & \textbf{-1.716}$^{*}$ & \textbf{-0.228}\\
        & [-1.822] & [-0.309]\\
        \hline
        Observations & 167 & 126\\
        R-squared & 0.050 & 0.084\\
        \hline
        \hline
    \end{tabular}
    \end{adjustbox}
\caption{\textbf{Determinants of CARs - Incident year}}\bigskip
\footnotesize{Results of regressions of the CARs on years. t-statistics are reported in brackets. The CARs are multiplied by 100. The first column uses the CARs obtained using OLS regressions, and the other column uses the CARs obtained with SUR. *, **, and *** indicate significance at the 10\%, 5\% and 1\% levels, respectively.}
\label{tab:CAR_on_years}
\end{table}

We cannot identify any monotonic increase or decrease of CARs magnitude through time, as previously advocated by \textit{e.g.} Gordon, Loeb, and Zhou, who find a decrease in the effect in their sample \cite{GordonLoebZhou2011}. Moreover, in the OLS specification (with 167 observations), we find that the coefficient for 2013 is positive, whereas, for the SUR estimation, that of 2020 is positive, albeit not statistically significant. Moreover, only two yearly dummy coefficients are statistically significant at the 10\% level (2016 for SUR and 2022 for OLS), and only one is significant at the 5\% level (2018 for SUR). Finally, the resulting $R^2$ is small in both estimations, with 5\% for the OLS case and 8.4\% for the SUR. These additional pieces of evidence make us confident that no time structure is present in the effects of cyber incidents on firms' valuation. Thus, we are not able to reject our null hypothesis $H_3$.

\subsection{Event type and target sector}

To test the hypotheses $H_{4a}$ and $H_{4b}$, we adopt the same panel regression setting and regress CARs on the type of cyber incidents reported in the news headlines, the sector the firms belong to, and their interactions. We report our results in Table \ref{tab:CAR_on_sector_type}.

\newgeometry{top=1in}

\begin{table}[H]
    \begin{adjustbox}{width= 0.50\textwidth,center}
    \begin{tabular}{lcccc}
        \hline
        \hline
        \multicolumn{5}{c}{Dependent variable: CAR}\\
        \hline
        & \multicolumn{2}{c}{OLS} & \multicolumn{2}{c}{SUR}\\
        \cmidrule(lr){2-3}
        \cmidrule(lr){4-5}
        & Model 1 & Model 2 & Model 1 & Model 2\\
        \hline
        Data breach & \textbf{-1.287}$^{*}$ & \textbf{0.261} & \textbf{-1.178}$^{**}$ & \textbf{-1.220}\\
        & [-1.907] & [0.167] & [-2.106] & [-1.013]\\
        Software breach & \textbf{-0.380} & & \textbf{-0.178}\\
        & [-0.480] & & [-0.290]\\
        Cyber breach & \textbf{1.152} & & \textbf{1.270}\\
        & [0.438] & & [0.349]\\
        Social breach & \textbf{0.081} & & \textbf{-0.086}\\
        & [0.045] & & [-0.061]\\
        Ransomware & \textbf{0.395} & & \textbf{-0.851}\\
        & [0.258] & & [-0.705]\\
        Shutdown & \textbf{-0.436} & & \textbf{-0.753}\\
        & [-0.041] & & [-0.747]\\
        Twitter breach & \textbf{-0.846} & & \textbf{-0.683}\\
        & [-0.070] & & [-0.817]\\
        Facebook breach & \textbf{2.378} & & \textbf{1.786}\\
        & [1.010] & & [1.096]\\
        Stolen funds & \textbf{0.362} & & \textbf{-0.617}\\
        & [0.135] & & [-0.231]\\
        Mitigation  & \textbf{1.762} & & \textbf{1.764}\\
        & [0.748] & & [1.083]\\
        Other & & \textbf{1.362} & & \textbf{0.422}\\
        & & [0.960] & & [0.360]\\
        \\
        Technology & & \textbf{-2.172} & & \textbf{-0.923}\\
        & & [-1.413] & & [-0.732]\\
        Consumer products & & \textbf{1.561} & & \textbf{-0.272}\\
        & & [0.793] & & [-0.166]\\
        Financials & & \textbf{-3.825} & & \textbf{0.955}\\
        & & [-0.722] & & [0.856]\\
        Healthcare & & \textbf{-5.210} & & \textbf{1.399}\\
        & & [-1.594] & & [0.500]\\
        Industrials & & \textbf{-1.713} & & \textbf{-3.471}\\
        & & [-0.680] & & [-1.455]\\
        \\
        Data breach * Technology & & \textbf{0.434} & & \textbf{0.329}\\
        & & [0.285] & & [0.270]\\
        Data breach * Consumer products & & \textbf{-2.341} & & \textbf{0.599}\\
        & & [-1.173] & & [0.380]\\
        Data breach * Financials & & \textbf{3.799} & & \textbf{0.955}\\
        & & [0.799]  & & [0.856]\\
        Data breach * Healthcare & & \textbf{-0.251} & & \textbf{-7.069}$^{*}$\\
        & & [-0.080] & & [-1.917]\\
        Data breach * Industrials & & \textbf{-1.381} & & \textbf{3.966}$^{*}$\\
        & & [-0.524] & & [1.708]\\
        \hline
        Observations & 167 & 167 & 126 & 126\\
        R-squared & 0.042 & 0.106 & 0.085 & 0.109\\
        \hline
        \hline
    \end{tabular}
    \end{adjustbox}
\caption{\textbf{Determinants of CARs - Incident type and target sector}}\bigskip
\footnotesize{Results of regressions of the CARs on the firm's sector and the type of the incident. t-statistics are reported in brackets. The CARs are multiplied by 100. \quotes{Consumer products} regroups the consumer cyclical and consumer non-cyclical sectors. The first two columns use the CARs obtained using OLS regressions, and the other two columns use the CARs obtained with SUR. *, **, and *** indicate significance at the 10\%, 5\% and 1\% levels, respectively.}
\label{tab:CAR_on_sector_type}
\end{table}

\restoregeometry
\onehalfspacing



In Model 1, we report the results of CARs regressed in a panel on dummy vectors set to \quotes{1} (\quotes{0}) when the incident belongs (does not belong) to the type of cyber incidents under scrutiny. The most significant incident affecting firm returns is the data breach, with an economic magnitude of -1.287\% (-1.178\%) with CARs estimated with the OLS (SUR) approach. Over the full sample, it translates into total, average, and median losses of USD 129 billion, 1.9 billion, and 105 million, respectively. The data breach coefficient is also the single one to be statistically significant at the 10\% level (5\% level) for CARs estimated with OLS (SUR). Not only do none of the other events seem to explain the CARs statistically, but they also show interesting coefficient sizes. For instance, the \quotes{cyber breach}, \quotes{social breach}, \quotes{ransomware}, \quotes{Facebook breach}, \quotes{stolen funds}, \quotes{mitigation}, as well as the \quotes{other} categories, all show positive signs for their coefficient. This coefficient magnitude reaches even 2.378\% for the \quotes{Facebook breach} type of incident (\textit{i.e.} when a company's Facebook account has been compromised) for CARs estimated with OLS.
In Model 2, we use the same approach and regress CARs in a panel on dummy vectors coded \quotes{1} when the firm belongs to a specific sector (technology, consumer products, financials, healthcare, and industrials) as well as their interaction with the data breach dummy vector. Given the few remaining degrees of freedom, we restrict the interaction terms with this type of cyber incident because the data breach coefficient is the only statistically significant in the Model 1 specification. We find that none of the sector coefficients is statistically significant, with the healthcare sector coefficient being the most negative at -5.21\% and approaching the 10\% statistical significance level for the CARs arising from the OLS estimation.\footnote{In an unreported test, we also regress CARs only on the sectors, dropping the interaction terms. The coefficients are virtually the same, except for the healthcare sector coefficient, which reaches the usual statistical levels. These results are available upon request} Over the full sample, it translates into total, average, and median losses of USD 12 billion, 1.2 billion, and 65 million, respectively. 

The financials sector ranks second with -3.83\%, followed by technology (-2.17\%) and industrials (-1.71\%). Conversely, the consumer products sector coefficient is positive at 1.56\%. Model 2, estimated with CARs estimated with the SUR approach, shows a different pattern. First, the sign of three sector dummy vector switches. In particular, that of the healthcare sector becomes positive. However, this is largely because the effect is captured by the interaction with the data breach dummy vector, which reaches -7.07\% and is statistically significant at the 10\% level. This supports the view that the marginal effect of having a data breach for a healthcare company is the most detrimental situation a company can bear around a cyber incident. The information content does not allow us to conclude why this specific combination is this detrimental. However, we can hypothesize that investors' view about potentially compromised private clinical data is particularly affected. This specification also yields the highest $R^2$ close to 11\%. Given the absolute contribution of the data breach effect versus the other types of incidents and its relative effects when combined with the healthcare sector, we find support to reject our null hypotheses $H_{4a}$ and $H_{4b}$. These results also align with those of Gordon, Loeb, and Zhou, who identify that data breaches with availability concerns are the most prone to generate negative abnormal returns \cite{GordonLoebZhou2011}.

\subsection{Firm characteristics}

We now study the firm characteristics as explanatory variables to test our hypothesis $H_{4c}$. We repeat the panel data regression using continuous variables for those characteristics. We report the results in Table \ref{tab:CAR_on_characteristics}.\footnote{We only report the OLS for the full sample (150 observations) and the SUR estimation (117 observations) for the restricted sample. The small statistical significance observed for the constant and price-to-earnings ratio only comes from the observation restrictions. In unreported OLS tests on the restricted sample, we also observe this significance arising. These results are available upon request.}

\begin{table}[H]
    \begin{adjustbox}{width= 0.38\textwidth,center}
    \begin{tabular}{lcc}
        \hline
        \hline
        \multicolumn{3}{c}{Dependent variable: CAR}\\
        \hline
        & OLS & SUR \\
        \cmidrule(lr){2-2}
        \cmidrule(lr){3-3}
        & Model 5 & Model 5\\
        \hline
        Constant & \textbf{-6.040} & \textbf{-7.872}$^*$\\
        & [-1.177] & [-1.682] \\
        Firm size (ln) & \textbf{0.206} & \textbf{0.348}$^*$\\
        & [0.994] & [1.881] \\
        Firm Age (ln) & \textbf{0.155} & \textbf{-0.414}\\
        & [0.741] & [-0.946] \\
        Book to market & \textbf{-0.065} & \textbf{-0.035}\\
        & [-0.644] & [-0.247] \\
        Price to earnings & \textbf{0.001} & \textbf{-0.015}$^{**}$\\
        & [0.115] & [-2.164] \\
        \hline
        Observations & 150 & 117\\
        R-squared & 0.017 & 0.075\\
        \hline
        \hline
    \end{tabular}
    \end{adjustbox}
\caption{\textbf{Determinants of CARs - Firm characteristics}}\bigskip
\footnotesize{Results of regressions of the CARs on firm characteristics. t-statistics are reported in brackets. The CARs are multiplied by 100. The first column uses the CARs obtained using OLS regressions, and the other column uses the CARs obtained with SUR. Certain observations were dropped due to missing accounting data. *, **, and *** indicate significance at the 10\%, 5\% and 1\% levels, respectively.}
\label{tab:CAR_on_characteristics}
\end{table}

The explanatory power of the considered characteristics is limited, with 1.7\% for the CARs estimated with OLS and 7.5\% for those estimated with the SUR approach. Only two variables of interest are statistically significant at the 5\% level (price-to-earnings ratio) and the 10\% level (firm size) in the SUR specification. To summarize these results, the larger the market capitalization of a firm but the smaller its price-to-earnings ratio, the more resilient it would be to cyber incidents. Nonetheless, most variation is unexplained and captured by the constant at -6.04\% and -7.87\% in the OLS and SUR specifications, respectively.

\subsection{News source effect}

To test our hypothesis $H_5$, we regress CARs on the type of news source to test whether this influences the economic magnitude and statistical significance of the costs of cyber incidents. We report our results in Table \ref{tab:CAR_on_source}.

\begin{table}[H]
    \begin{adjustbox}{width= 0.30\textwidth,center}
    \begin{tabular}{lcc}
        \hline
        \hline
        \multicolumn{3}{c}{Dependent variable: CAR}\\
        \hline
        & OLS & SUR \\
        \cmidrule(lr){2-2}
        \cmidrule(lr){3-3}
        & Model 4 & Model 4\\
        \hline
        Reuters & \textbf{0.186} & \textbf{-0.477}\\
        & [0.219] & [-0.718] \\
        Twitter & \textbf{-0.843} & \textbf{-1.029}\\
        & [-0.313] & [-0.549] \\
        Other & \textbf{-0.840} & \textbf{-0.472}\\
        & [-1.239] & [-0.927]\\        
        \hline
        Observations & 167 & 126\\
        R-squared & 0.001 & 0.006\\
        \hline
        \hline
    \end{tabular}
    \end{adjustbox}
\caption{\textbf{Determinants of CARs - News source}}\bigskip
\footnotesize{Results of regressions of the CARs on the news source. t-statistics are reported in brackets. The CARs are multiplied by 100. The first column uses the CARs obtained using OLS regressions, and the other column uses the CARs obtained with SUR.}
\label{tab:CAR_on_source}
\end{table}

We have 270 news on the unrestricted sample, from which we can estimate CARs with OLS, which is split between 156 news provided by Reuters, 15 by Twitter, and the rest classified as \quotes{other}. Different source types are extremely diverse and generally count as one. The split for the restricted sample of 126 CARs is 72 for Reuters and 4 for Twitter, the remainder classified as \quotes{other}. We find almost no explanatory power with these panel regressions, with an $R^2$ at 0.1\% for CARs arising from OLS and 0.6\% for those deriving from the SUR estimation. Moreover, we cannot find any significant coefficients at the usual statistical levels for the type of news source. Interestingly, the coefficient for Reuters is positive when CARs from OLS are regressed. In contrast, the most considerable negative magnitude is found for news from Twitter for CARs from both estimations, with -0.84\% and -1.03\% for CARs from OLS and SUR estimations, respectively. Thus, we cannot reject our null hypothesis $H_5$, which supports the view that the source of the first available news does not affect CARs' magnitude.

\subsection{Robustness tests}

We repeat the core analysis with a zero-benchmark model to ensure a misspecified benchmark does not drive our results. We present these results in Table \ref{tab:average_car_tstat}, Figure \ref{fig:AAR_zb}, and Figure \ref{fig:CAAR_zb}. The results and significance are very similar to the ones obtained with the three-factor benchmark, which makes us confident that our benchmarks do not generate bias. An alternative way to consider these results is that the procedure generates the actual \quotes{raw} return that an average agent would experience while being invested in these stocks over the cyber incident windows.

%% file: sections/discussion.tex





\subsection{Hypotheses explaining our results}

Several factors could explain our findings regarding the health sector and the data breach. However, these explanations are hypothetical at this stage and can only be based on the patterns observed in our results.
First, data breaches involve unauthorized access to sensitive information, often leading to compromised personal or confidential data. The health sector deals with highly sensitive patient information, making data breaches particularly damaging. The potential for compromised clinical data, patient records, or personal health information may severely affect individuals and healthcare organizations. Second, the health sector is subject to strong regulations and compliance standards, such as the US's 1996 Health Insurance Portability and Accountability Act (HIPAA).\footnote{See, \url{https://www.govinfo.gov/content/pkg/CRPT-104hrpt736/pdf/CRPT-104hrpt736.pdf}} The regulatory environment demands a high level of data protection. Any failure to safeguard patient information can result in significant legal and financial consequences. This regulatory scrutiny amplifies the impact of data breaches in the health sector compared to other industries, which may also explain why the combined effect of data breaches on the health sector's firms is even more detrimental. Third, health data is often considered more valuable than other information on the dark web due to its potential for various malicious activities, including identity theft and healthcare fraud. As a result, cybercriminals may specifically target the health sector to gain access to valuable data, contributing to the sector's increased vulnerability. Finally, data breaches involving sensitive patient information can significantly erode public trust, leading to a permanent price depreciation (negative CARs). The potential harm to an organization's reputation and the loss of patient confidence can have lasting effects, influencing investors' perceptions and contributing to a more pronounced negative impact on market valuation.
In contrast, the non-significant CARs we observe on ransomware attacks may be explained by their impact being more immediate and operational rather than directly affecting a firm's valuation. Organizations that have robust backup and recovery mechanisms may be able to mitigate the financial losses associated with such attacks. Additionally, the financial demands of ransomware may not necessarily translate into a direct and lasting impact on a firm's market valuation, explaining the observed lesser significance.

It is important to note that these explanations are speculative. Further research would be needed to validate and refine these hypotheses, considering additional factors and potential interactions that may influence the impact of cyber incidents on different types of firms and sectors.

\subsection{Policy recommendations}

Given the significant negative impact on firms in the health sector, policymakers should consider sector-specific cybersecurity guidelines and incentives to mitigate the vulnerabilities identified in the study. This could include tailored regulations and support for health-related organizations to enhance their cybersecurity posture.

Policymakers should encourage and enforce robust data protection measures, possibly through sector-specific regulations addressing the unique challenges of data breaches. They should also promote mechanisms for improved information sharing among firms within the same sector. We observe that firms from all sectors suffer from data breaches, and thus, encouraging collaboration and sharing best practices may enhance firms' overall cybersecurity resilience. Next, since we find that cyber incidents do not uniformly impact all firms, policymakers should work with the insurance industry to develop contracts mitigating the impact of each specific cyber incident. This could involve tailoring insurance coverage and premiums based on the sector, type of cyber incident, and other relevant factors identified in the study. 

Last, there should be increased collaboration between government agencies, private organizations, and cybersecurity experts to address cyber incidents' challenges. The policy approach should be dynamic and targeted, considering the nuanced nature of cyber threats. Policymakers should address specific vulnerabilities identified in the study and collaborate with industry stakeholders to comprehensively respond to the evolving nature of cyber incidents.

\subsection{Cybersecurity investment recommendations}

Given the significant negative impact identified for firms in the health sector, there is a strong case for these organizations to contract cyber insurance products. Insurance policies should be tailored to cover potential economic losses associated with data breaches.
While our results are not statistically significant for firms in sectors other than healthcare, it is still advisable for all firms to consider cyber insurance. The coverage should be structured to align with each sector's specific risks. In this case, standard cyber insurance policies may be sufficient for firms outside the health sector. Firms in the health sector should prioritize investments in cybersecurity measures, especially those focused on preventing and mitigating data breaches. This may include implementing advanced encryption technologies, access controls, and employee cybersecurity training to prevent breaches. Additionally, allocating resources for continuous monitoring and threat intelligence can enhance the ability to promptly detect and respond to cyber threats.

Overall, our results point to the specificities of the impact of cyber incidents, and as such, investment should be tailored to the specific risks of each sector. Cybersecurity investments and insurance policies should be strategic and customized to cover these unique risks faced by each sector.

%% file: sections/conclusion.tex
Our study delves into the impact of cyber incidents on listed firms, employing advanced event study methods and adjusted returns. By utilizing newswire headlines filtered for relevant information on cyber incidents, we use state-of-the-art estimation methods, specifically a seemingly unrelated regression estimation, to ensure unbiased cumulative abnormal returns coefficients. Our findings reveal an overall negative effect of approximately -0.89\%, regardless of the target firm or the type of cyber incident.

We adjust standard errors using corrections for event-induced variance and cross-correlation to address a crucial caveat in the existing literature. Unadjusted standard errors result in cumulative abnormal returns with significance below the 1\% level, but our correction reveals no statistical significance at the aggregated level. This contradicts the prevailing consensus in event study literature. A more detailed analysis incorporating variables such as the type of cyber incident and sector of the target firm reveals a significant marginal effect on firms in the health sector and those involved in data breaches. In contrast, counterintuitively, ransomware or other attacks do not significantly affect firms' value. No heightened sensitivity is observed in the financial sector. Conversely, alternative explanatory variables such as time-fixed effects, firm characteristics, or news sources yield little to no explanatory power and statistical significance.

One alternative benchmark to be considered would be the \quotes{peer firms benchmark}, \textit{i.e.} building the counterfactual returns using a set of firms with similar characteristics. This would help reduce the systematic pre-event systematic positive returns we observe. However, we would be limited in our setting as firms with characteristics identical to those subject to cyber incidents would also be more frequently affected. Another possible extension that would allow the research to be more conclusive on the permanent \textit{transitory} price pressure debate is to study the effects of cyber incidents on returns and other market metrics such as turnover, volatility, liquidity (bid-ask spread), etc.

By adopting a standard error correction approach, an up-to-date dataset reflecting the changes in cyber incident frequency, and a new set of headlines filtered with NLP methods, we argue that cyber incidents have less systematic detrimental effects on firms' value than previously claimed when considered aggregated. However, we identify some edge cases, mainly when cyber incidents are data breaches and when target firms belong to the health sector, which is particularly detrimental for the firm. Our results also highlight the heterogeneity of the impact of cyber incidents, and not only calls for improved investments and insurance contracts for the health sector regarding their data but also for taking into account each specific risk across each sector.

%% file: sections/appendix.tex
\phantom{test}
\vspace{60mm}
\begin{table}[h]
\begin{adjustbox}{width=\textwidth,center}
    \begin{tabular}{lcc}
    \hline
    \hline
     Variable & Description & Source \\
    \hline
    Firm size (ln) & ln(market equity [prc*shrout])  & CRSP\\
    Firm Age (ln) & ln(years) since the firm first appeared in Compustat & Compustat\\
    Book to market ratio & Common equity [ceq] / market equity & Compustat and CRSP\\
    Price/Earnings & Stock Price / Earnings [pe\_exi] & WRDS Financial Ratios\\
    \hline
    \hline
    \end{tabular}
    \end{adjustbox}
    \caption{\textbf{Variable definitions}}\bigskip
    \footnotesize{The names of the variables as found on CRSP and Compustat are in brackets.}
\label{tab:variable_descriptions}
\end{table}

\clearpage
\begin{figure}
    \noindent\makebox[\textwidth]{%
    \includegraphics[width=\textwidth]{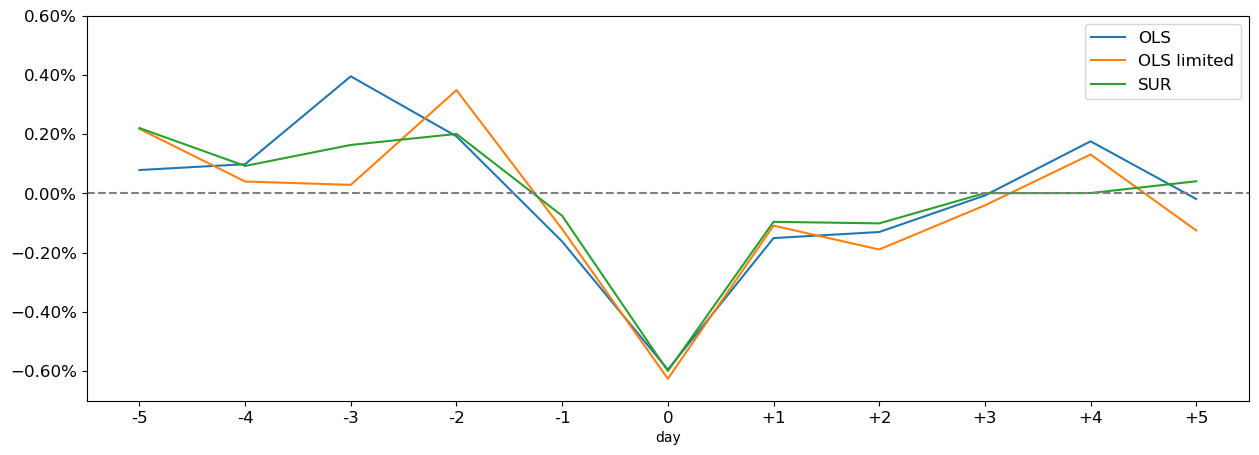}
    }
     \caption{\textbf{Average abnormal returns - Zero benchmark}}\bigskip
     \footnotesize{Average abnormal daily returns in an 11-day window centered on the cyber events. Day 0 is the day of the cyber event. The OLS limited model is the OLS model restricted to the same period, the same subset of firms and events as the SUR model. Abnormal returns are computed with respect to the zero-benchmark model, following Eq. \ref{eq:AR_regression_zb}.}
     \label{fig:AAR_zb}
\end{figure}

\clearpage
\begin{figure}
    \noindent\makebox[\textwidth]{%
    \includegraphics[width=\textwidth]{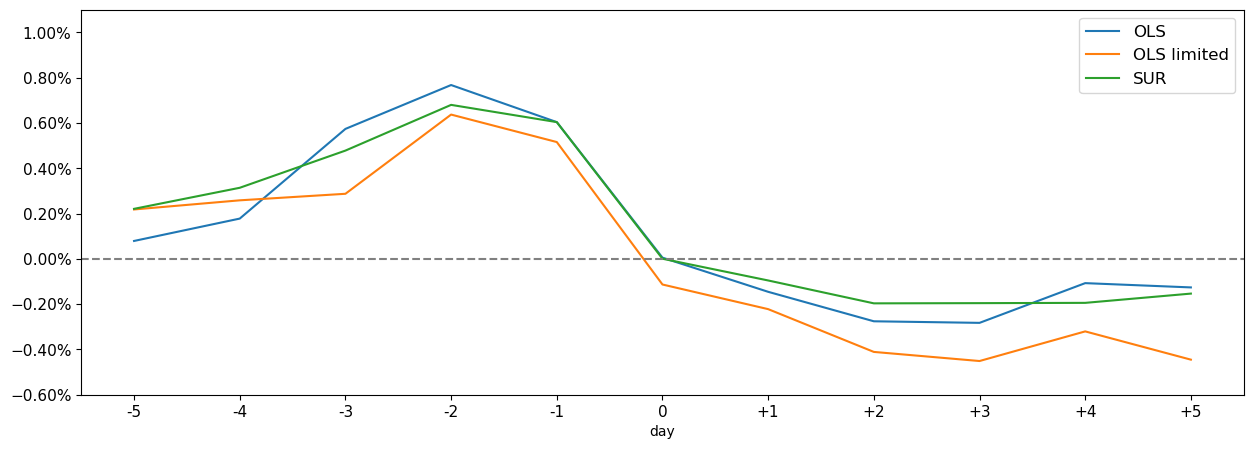}
    }
     \caption{\textbf{Cumulative average abnormal returns - Zero benchmark}}\bigskip
     \footnotesize{Cumulative average abnormal daily returns in an 11-day window centered on the cyber events. Day 0 is the day of the cyber event. The OLS limited model is the OLS model restricted to the same period, the same subset of firms and events as the SUR model. Abnormal returns are computed with respect to the zero-benchmark model, following Eq. \ref{eq:AR_regression_zb}.}
     \label{fig:CAAR_zb}
\end{figure}